# Thermoelastic properties and thermal evolution of the Martian core from ab initio calculated ferromagnetic Fe-S liquid

Wei-Jie Li[1,3], Zi Li[2,4], Zhe Ma[1,3], Jie Zhou[1,3], Cong Wang[2,4,5], Ping Zhang[2,4,5]

[1] Intelligent Science & Technology Academy Limited of CASIC, Beijing, 100141, People's Republic of China

[2] Institute of Applied Physics and Computational Mathematics, Beijing, 100088, People's Republic of China

[3] Scientific Research Key Laboratory of Aerospace Defence Intelligent Systems and Technology, Beijing, 100141, People's Republic of China

[4] Tianfu Innovation Energy Establishment, Chengdu, 610213, China

[5] Center for Applied Physics and Technology, Peking University, Beijing, 100871, People's Republic of China

**Abstract**

The accurate thermoelastic properties and thermal conductivity are crucial in understanding the thermal evolution of the Martian core. A fitting method based on the ab initio calculated pressure-volume-temperature data is proposed in the formulation of the equation of state with high accuracy, by which the pressure and temperature dependent thermoelastic properties can be directly calculated by definitions. The ab initio results show that the liquid $Fe_{0.75}S_{0.25}$ under Martian core condition is thoroughly in the ferromagnetic state, without existing spin crossover. The liquid $Fe_{0.75}S_{0.25}$ in magnetic calculation owns a low thermal conductivity (21~23 W/m/K) when compared with non-magnetic calculation at the same state. Based on the Insight estimated and ab initio calculated properties of the Martian core, the iron snow model is verified when the current temperature at the core-mantle boundary is below the core melting temperature, and the simply secular cooling model is verified on the contrary.

## 1. Introduction

The study of the Martian core is critical for our general understanding of planetary evolution. The recent Martian explorations provide direct constraints for the Martian interior structure of Mars [1-4]. Mars possessed a short-lived (~0.5 billion years) early Martian core dynamo and does not possess a global magnetic field today, but the regions of its crust are strongly magnetized. The basic property constraint the evolution of Mars, which gives implications for an early dynamo [5]. One primary motivation for the present study is the anticipation of the deep interior of the Martian core which is still uncertain, including the equation of state, thermoelastic properties, thermal conductivity, and the possible thermal evolution scenarios.

A variety of scenarios for the evolution of Mars have been proposed, involving different physical processes. The inner core or iron snow zone exists when phase transformation appears in core [6]. When the intersection ($R_s$) between the core adiabatic temperature curve and the composition-dependent melting curve is on top of the core, it is iron-snow model [7,8]. When the intersection $R_s$ is on the bottom of the core, it is solid iron inner or sulfide inner core [7]. The stratification exists on top of the core when heat flux is sub-adiabatic [9] or immiscibility of Fe-S-H liquid [10] exists. Without no phase transformation or stratification, it is a fully liquid core with thermal convection and simply secular cooling. By calculating the contributions of the physical processes to the energy and entropy budgets [8,11], the thermal evolution of the Martian core and its dynamo are collected. The dynamo in the Martian core is influenced by the sulfur concentration, thermoelastic properties, the thermal conductivity of the core, and the heat flow out of the core. The heat flows out of the core is closely related to the properties of the Martian mantle, which is not the main concern in this paper.

The thermal evolution of the Martian core is sensitive to the intersection ($R_s$) between the core adiabatic temperature curve and the composition-dependent melting curve. The slope ($dT/dP$) of the melting line of the Fe-S system with (S<16 wt.%) is negative when S concentrations are 10.6 wt.% and 14.2 wt.%, and positive when S concentration is 16.2 wt. % [7]. The melting line of the FeNi-S system indicated that the core crystallizes as iron snow with 10-13 wt.% and crystallizes as $Fe_3S$ at the center with 15-16 wt.% [6]. The thermoelastic properties and core adiabatic temperature are calculated by the equation of state of the Fe-S system, which was described by an ideal solution model [12,13]. By the ultrasonic pulse-echo overlap method combined with a Kawai-type multi-anvil apparatus experiments, the sound velocity of Fe-S liquid under Martian core conditions is extrapolated and is least sensitive to S concentration in the whole Martian pressure range [14]. The iron-sulfide equation of state is crucial for modeling the internal structures, which need to be accurately constructed.

In the thermal evolution, thermal stratification and cessation time of dynamo are sensitive to thermal conductivity [15]. Electrical resistivity measurements on hcp Fe alloy containing 3 wt.% S shows that the thermal conductivity of the Martian core range from ~46-60 W/m/K [16]. The thermal conductivity of pure liquid Fe under Martian core condition is ~82±3 W/m/K by current reversal technique experiments on the resistivity of Fe, and an increase in thermal conductivity from low pressure to the CMB of Martian core ascribed to the loss in long range order magnetic structure [17]. The thermal conductivity of Fe-Si solid under Martian core conditions is ~36 W/m/K by ultrafast time-domain thermoreflectance with diamond anvil cell technique, and Si addition decreases the thermal conductivity of Fe [18]. The S additional effect on the magnetic structure and thermal conductivity of liquid Fe under Martian core conditions is unexplored.

Ab initio calculations were adopted to calculate the magnetic structure, equation of state, thermoelastic property, and thermal conductivity of Fe-S fluid under Martian core conditions. Firstly, the equation of state is fitted by the ab initio calculated pressure-volume-temperature data and considering the magnetic structure with 1% accuracy. After establishing the equation of state, the thermoelastic properties at Martian core conditions are collected directly by its definitions. Secondly, the thermal conductivity of Fe-S fluid was collected at possible Martian core conditions with both spin-polarized and spin-non-polarized calculations. At last, the thermoelastic properties and thermal conductivity along the adiabatic curve were adopted in the thermal evolution of the Martian core.

## 2. Methods and calculations

To model the thermal evolution of the Martian core, possible mechanisms should be established. To calculate these thermal evolution processes, the basic property of the Martian core should be collected. Then, the basic ab initio simulation methods and parameters are illustrated. From the ab initio calculated results, the basic properties are derived, and possible thermal evolution is concluded.

### 2.1. Thermal evolution of Martian core

A spherically averaged, one-dimensional thermodynamic model of the core depending only on the radius is assumed. The basic properties of the Martian core can be calculated by the thermoelastic properties [12,13]. The gravity acceleration g(r) obeys the Poisson equation [13]

$$\frac{dg(r)}{dr} + \frac{2g(r)}{r} = 4\pi G \rho(r) \tag{1}$$

where $r$ is the radial distance from the Martian core, and $G$ is the gravitational constant. The core density $\rho(r)$ is

$$\frac{d\rho(r)}{dr} = -\frac{-\rho^2(r) g(r)}{K_s(r)} \tag{2}$$

where $K_s$ is the adiabatic bulk modulus.

For a well-mixed and vigorously convecting core, the temperature $T_{ad}(r)$ is

$$\frac{\partial T_{ad}(r)}{\partial r} = -\frac{\gamma(r)}{K_s(r)} \rho(r) g(r) T_{ad}(r) \tag{3}$$

where $\gamma$ is the Grüneisen parameter.

The hydrostatic pressure $P(r)$

$$\frac{dP(r)}{dr} = -\rho(r) g(r) \tag{4}$$

Then the radial distribution of $\rho$, $g$, $T_{ad}$, and $P$ are collected by Equation (1)-(4) with the boundary condition that $g(0)=0$, $T_{ad}(r_{cmb})=T_{cmb}$, and $P(r_{cmb})=P_{cmb}$, where $T_{cmb}$ and $P_{cmb}$ are temperature and pressure at CMB ($r_{cmb}$), respectively. The basic properties of the Martian core are based on recently reported Insight data [3]: the radius of the core is 1830±40 km, the mean core density is 5.7-6.3 g/cm$^3$, the adiabatic temperature at CMB is ~1900-2000 K, the adiabatic pressure is 18-19 GPa, and S content is 10-20 wt.% [12,13].

The primary mechanisms for driving core convection are generally thought to be thermal (due to rapid core-mantle heat flow) and compositional. We assume that there are no radioactive elements in the Martian core. In this calculation, we just consider secular cooling, phase transformation on top (top-down mechanism, iron snow model [8]), and bottom (bottom-up mechanism, nucleation of the inner core, similar to Earth) of the core. If the intersection of the adiabatic curve and melting line is on top of the core, it is the up-bottom mechanism (iron snow model). If the intersection ($R_s$) of the adiabatic curve and melting line is on the bottom of the core, it is the inner core of iron as in Earth's core or maybe Fe$_3$S inner core.

The energy budget of the Martian core is

$$Q_{cmb} = Q_s + Q_g + Q_L = \tilde{Q}\frac{dT_{cmb}}{dt} \tag{5}$$

where $Q_{cmb}$ is the heat flux at the core-mantle boundary (CMB), $Q_s$ is the secular cooling of the core, $Q_g$ is the gravitational energy associated with the separation of crystallization and $Q_L$ is the latent heat related to the crystallization. As the $Q_{cmb}$ is closely correlated with mantle properties, we adopted a parameterized model of the time evolution of $Q_{cmb}(t)$ for simply [5]. The detailed formulations in Eq.(5) are illustrated in [8,11]. If there is no phase transformation or no intersections between the adiabatic curve and melting, $Q_g=Q_L=0$, the core is simply secular cooling. From the Eq. (5), the time evolution of $T_{cmb}(t)$ is calculated.

The entropy budget to determine the dynamo is

$$E_J + E_k = E_s + E_g + E_L = \tilde{E}\frac{dT_{cmb}}{dt} \tag{6}$$

where $E_J$, $E_k$, $E_s$, $E_g$, and $E_L$ are entropies from Ohmic dissipation, heat conduction entropy, secular cooling, gravitational energy, and latent heat of crystallization, respectively. $E_k$ is calculated from thermal conductivity $k$, $E_k = \int k\left(\frac{\nabla T}{T}\right)^2 dV$. The detailed formulations in Eq.(6) are illustrated in [8,11]. From the calculated $T_{cmb}(t)$, the thermal evolution of $E_J(t)$ is calculated from Eq. (6). $E_J$ is used to determine when the dynamo of Martian may be active.

### 2.2. Equation of state and thermal conductivity

In this paper, the pressure-volume-temperature (P-V-T) and energy-volume-temperature (E-V-T) data are fitted by multivariate polynomial method

$$\begin{aligned} P &= \sum_{i,j} A_{ij} V^i T^j \\ E &= \sum_{i,j} B_{ij} V^i T^j \end{aligned} \tag{7}$$

where $V$ is the volume per atom. From the fitted equation of state of Fe-S liquid under Martian core conditions, the thermoelastic properties at the specified state are collected by its definitions [19], such as isothermal bulk modulus ($K_T$), adiabatic bulk modulus ($K_s$), thermal expansion coefficient ($\alpha$), Grüneisen constant ($\gamma$), heat capacity at constant volume ($C_V$) and sound velocity ($V_p$).

The Kubo-Greenwood formula for the electrical conductivity as a function of the frequency ω for a particular $k$ point in the Brillouin zone of the simulation supercell is written as

$$\sigma_{\mathbf{k}}(\omega) = \frac{2\pi e^2 \hbar^2}{3m^2 \omega \Omega} \sum_{j=1}^{N} \sum_{i=1}^{N} \sum_{\alpha=1}^{3} \left[ F(\varepsilon_{i,\mathbf{k}}) - F(\varepsilon_{j,\mathbf{k}}) \right]$$
$$\times \left| \langle \Psi_{j,\mathbf{k}} | \nabla_\alpha | \Psi_{i,\mathbf{k}} \rangle \right|^2 \delta(\varepsilon_{j,\mathbf{k}} - \varepsilon_{i,\mathbf{k}} - \hbar\omega) \quad (8)$$

The thermal conductivity ($k$, W/m/K) is calculated from the electronic current autocorrelation function via Kubo's linear response formalism

$$L_{ij} = (-1)^{i+j} \frac{he^2}{V} \sum_{l',l} \lim_{\varepsilon \to 0} \frac{f(\varepsilon_{l'}) - f(\varepsilon_l)}{\varepsilon} \delta(\varepsilon_{l'} - \varepsilon_l - \varepsilon)$$
$$\times \langle \psi_l | \mathbf{v} | \psi_{l'} \rangle \langle \psi_{l'} | \mathbf{v} | \psi_l \rangle (\varepsilon_{l'} - \varepsilon_f)^{i-1} (\varepsilon_{l'} - \varepsilon_f)^{j-1}, \quad (9)$$
$$k = \frac{1}{e^2 T} \left( L_{22} - \frac{L_{12}^2}{L_{11}} \right).$$

where $\varepsilon_f$ is the Fermi energy; $\psi_l$, $\varepsilon_l$, and $f(\varepsilon_l)$ are the wave function, eigenvalue, and Fermi-Dirac occupation of eigenstate $l$, respectively; $\mathbf{v}$ is the velocity operator; and $V$ is the simulation cell volume, $T$ is temperature. Thermal conductivity is correlated with electronic conductivity via Wiedemann-Franz law [20].

## 2.3. Calculation details

The *ab initio* molecular dynamics calculations are implemented in the plane wave density functional VASP code [21,22]. The thermal conductivity is calculated using the Kubo-Greenwood formula and the Chester-Thellung-Kubo-Greenwood formula as implemented in VASP. In this calculation, projector augmented waves (PAWs) [23,24] and generalized gradient approximation (GGA) in the parameterization of Perdew, Burke, and Ernzerhof [25] are adopted. The pseudopotential of Fe has p electrons as valence electrons and s electrons as standard electrons. Spin-polarized and spin-non-polarized calculations are adopted to get the equilibrium state and analyze the possible magnetic collapse or magnetic multiplet in the Martian core. The Hubbard $U$ model is implemented to improve the spin-polarized calculation. The U parameters were calculated by the linear response approach [26,27], and the U value of Fe-S liquid under Martian core condition is around 4.3 eV. The plane wave cutoff is 400 eV, which is sufficient to ensure that the pressure converges with 1% accuracy. The time-dependent mean square displacement is employed to check the system in the liquid state. The selected time step is 1 fs in all the calculations.

We selected 128 atoms as the cell. Atoms were randomly distributed in the cell, with 32 S atoms occupying random locations in the model, and the other atoms were Fe. The corresponding atom ratio of S is 16 wt.%, which is consistent with the recently reported Insight data [3]. The pressure ranges from 16 GPa to 40 GPa, and the temperatures range from 1600 K to 2600 K with 200 K as the step. To ensure the calculated pressure and temperature ranges, the *NPT* ensemble is calculated for 2000 time steps first, and then the equilibrium volume is computed under certain pressures and temperatures. Then, the *NVT* ensemble with the equilibrium volume is calculated with the total simulation time exceeding 6000 time steps. At last, the thermal conductivity is calculated from Eq. (9). The thermal conductivity ($k$) was obtained by averaging ten snapshots extracted from the last 4,000 time steps in each MD trajectory in the NVT ensemble with an interval of 400 time steps. We note that $k$ refers only to the electronic component.

## 3. Results and Discussions

### 3.1. Equation of state

From the ab initio results, Fe-S liquid is stable at the ferromagnetic state rather than at the nonmagnetic state, in which temperature and pressure ranges are 1600~2600 K and 16~40 GPa, respectively. The mean magnetic moments per atom of Fe-S liquid is almost a steady positive value. It shows that The Fe-S fluid is at the high-spin state (ferromagnetic) under the whole Martian core condition, which is different from pure solid iron experimental results (from ferromagnetic to paramagnetic state) [17] and Earth's low mantle study which also shows a complex magnetic state at 20 to 40 pressures [28]. This is also different from ref. [29], which consider a nonmagnetic $Fe_3S$ solid. The S elements addition makes sure that the Fe-S liquid at the Martian core condition is thoroughly ferromagnetic.

By hundreds of ab initio spin-polarized calculations, the pressure-volume-temperature and energy-volume-temperature data are collected. The pressure and energy are fitted by Eq.(7) to ensure convergence within 0.5 %, and the detailed fitting parameters are in Table 1. The multivariate polynomial is fitted by PolynomialFeatures in the sklearn package with the python programming language, which is enough to ensure the fitting convergence. The fitting standard error is ensured within 1%, which is the ab initio molecular dynamic calculated pressure and energy accuracy. With the formula expressions of P(V,T) and E(V,T), the thermoelastic properties can be directly derived by definitions.

Table 1 Fitting parameters of pressure and energy from temperature and volume in Fe-S liquid under Martian core condition. The detailed formula expressions are shown in Eq.(7).

| $A_{ij}$ | i=0 | i=1 | i=2 | $B_{ij}$ | i=0 | i=1 | i=2 |
|---|---|---|---|---|---|---|---|
| j=0 | 540.53 | -72.62 | 2.46 | j=0 | 1.58 | -0.992 | $3.54 \times 10^{-2}$ |
| j=1 | 0.0054 | $1.87 \times 10^{-7}$ | 0 | j=1 | $2.686 \times 10^{-4}$ | $5.053 \times 10^{-9}$ | 0 |
| j=2 | $-1.24 \times 10^{-4}$ | 0 | 0 | j=2 | $1.349 \times 10^{-6}$ | 0 | 0 |

### 3.2. Thermoelastic properties along the core adiabatic curve

From the equation of state, the temperature and pressure dependent thermoelastic properties under the whole Martian core condition are derived. Combing the Eq. (1)-(4) and ab initio calculated thermoelastic properties, the adiabatic pressure and temperature are calculated with different initial conditions. By selecting 1900K, 1600K, and 1300K as the current CMB temperature, the adiabatic pressure and temperature curve (Figure 1(a)), thermal expansion coefficient (Figure 1(b)), isothermal bulk modulus (Figure 1(c)), sound velocity (Figure 1(d)) and all the other thermoelastic properties along the adiabatic curve are collected. Temperature and pressure dependent isothermal bulk modulus and thermal expansion coefficient are of the same order as the early reports[11,26,27], which confirmed our calculated methods of thermoelastic properties.

The snow zone at the current time is directly determined by the intersections between the adiabatic curve and Fe-S liquid melting curve. Compared with the melting lines of Fe-S liquid (Figure 1(c)), the iron snow exists if the S content is 10.2 wt.% and $T_{cmb}$<1930 K at present. If the S content is higher than 16 wt.%, the present $T_{cmb}$ is higher than the melting line of Fe-16wt.%S, and a $Fe_3S$ inner core [7] may appear when the $T_{cmb}$ is sufficiently low in the future time and the iron inner core is impossible. If the S content is <16 wt.%, the beginning time of iron snow is correlated with the appearance of the intersection between the adiabatic curve and melting temperature. The possible thermal evolution mechanism in the Martian core at present may be iron snow or simply secular cooling, and $Fe_3S$ solid core may exist in the future. Detailed numerical calculations about the thermal evolution of the Martian core are shown in Section 3.4.

The P-wave sound velocity ranges from 4.21 to 5.19 km/s along the core adiabatic curve when present $T_{cmb}$ = 1900 K (Figure 1(d)). The calculated sound velocity is a little lower than the experimental data [14] as the calculated density of Fe-S liquid is higher than the reported average density of core [3]. The sound velocity is calculated by $V_P = \sqrt{K_s/\rho}$, which implies the sound velocity is low when the density is considerably high. Moreover, the S concentration and ferromagnetic state may both affect the thermoelastic properties, which is not discussed in this paper.

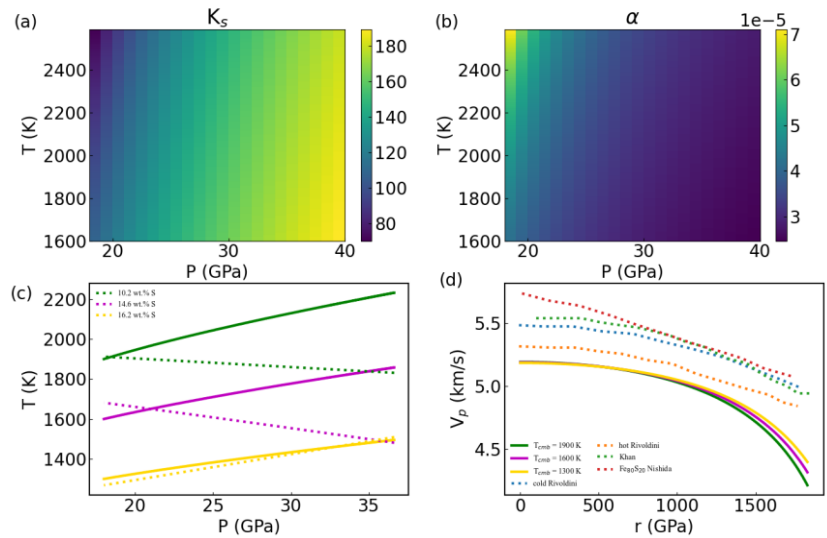

Figure 1 The calculated thermoelastic properties of $Fe_{0.75}S_{0.25}$ liquid under Martian core condition. (a) temperature and pressure dependent adiabatic bulk modulus $K_s(T,P)$, (b) temperature and pressure dependent thermal expansion coefficient $\alpha(T,P)$, (c) adiabatic pressure-temperature curve with different initial CMB temperature, and (d) P-wave sound velocity ($V_p$) along the adiabatic curve. The initial CMB temperatures are selected by 1900K, 1600K, and 1300K, respectively. The '10.2 wt.%S', '14.6 wt.%S', and '16.2 wt.%S' labels in subfigure (c) are the experimental melting lines of Fe-S liquid with different S concentrations [7]. The 'cold Rivoldini' [12], 'hot Rivoldini' [12], 'Khan' [13], and '$Fe_{80}S_{20}$ Nishida' [14] labels are the experimental compressional sound velocity.

## 3.3. Thermal and electronic conductivity

The thermal conductivities of Fe-S liquid in the ferromagnetic state (constant density curve in Figure 2) are lower than that in the nonmagnetic state (2000K isothermal curve in Figure 2). The thermal conductivity in the Martian core is in the ranges of 21~23 W/m/K at ferromagnetic state (in Figure 2), while in the ranges of 40~45 W/m/K at Martian core pressure scale at constant temperature 2000 K ('2000 K' label in Figure 2 (b)). Taking the magnetic state into consideration, the thermal conductivity at the ferromagnetic state is half the value of that at the nonmagnetic state.

The ab initio calculated thermal conductivities of Fe-S liquid at the nonmagnetic state are consistent with the experimental resistivity and thermal conductivity values, such as $Fe_{77.7}S_{22.3}$ [16], pure hcp Fe [17], and iron-silicon alloys [18], which are all double values of our calculated data at ferromagnetic state. The resistivity of S addition to hcp iron is lower than pure hcp iron and Si addition [16]. The result that thermal conductivity at the ferromagnetic state is lower than that at the nonmagnetic state is consistent with the pure iron experimental results [17]. The magnetic state affects the thermal conductivity values.

From the constant density curve at different temperatures in Figure 2, the pressure of 7130 kg/m$^3$ is higher than at 6130 kg/m$^3$. It is evident that density at the core is higher than that at the core-mantle boundary, but only within 10% density variation along the adiabatic curve. The density in the thermal evolution of the Martian core is set to be a constant. The low thermal conductivity corresponds to a low $Q_{ad}$, and the thermal stratification is unstable on top of the Martian core when $Q_{CMB} > Q_{ad}$. If there must exist a thermal stratification, the $Q_{CMB}$ should be lower, which also corresponds to a high order of mantle viscosity [9]. Thus, the thermal stratification near CMB is neglected in the thermal evolution calculation.

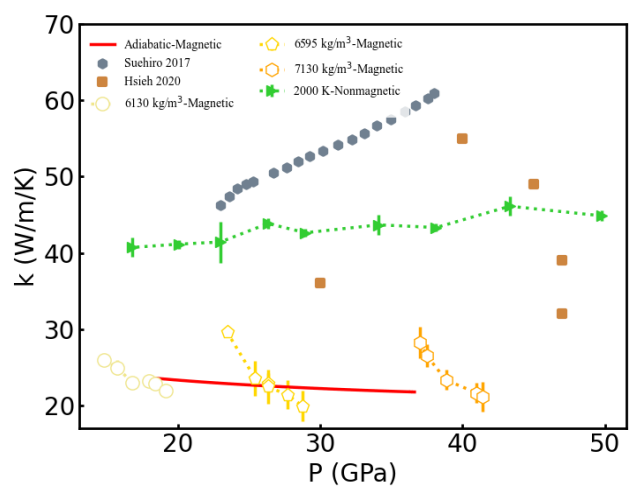

Figure 2 Ab initio calculated thermal conductivity of Fe-S liquid. Label '6130 kg/m$^3$', '6595 kg/m$^3$', and '7130 kg/m$^3$' are spin-polarized calculated data at constant volume. The '2000 K' label is ab initio calculated data at a constant temperature with spin-non-polarized calculation. "Suehiro 2017"

is experimental data of Fe-S alloy with experience equation from Ref. [16]. "Hsieh 2020" are experimental data of Fe-Si alloy from Ref. [18].

### 3.4. Thermal evolution of the Martian core

The possible mechanisms of thermal evolution are simply secular cooling, the iron snow model, the solid Fe inner core, and the $Fe_3S$ inner core. From Insight data about Martian core [3], the two possible thermal evolution mechanisms of the Martian core at present time are iron snow or simply secular cooling. From the calculated low thermal conductivity, the thermal stratification layer may not accumulate on top of the Martian core. A solid inner core of $Fe_3S$ may appear in the future if the center temperature of the core is below the melting temperature of $Fe_3S$ with an S concentration high enough. The phase diagram of the Fe-S system and the high S concentration show that an iron solid inner is impossible.

In the calculation, the present CMB temperature is selected as 1860 K to best describe the iron snow zone and the melting line is selected with 10.2 wt.% S. To explore the thermal evolution history of the Martian core, the current state is $P_{cmb}$ = 18 GPa and $\rho_c$= 6300 g/cm$^3$ (subfigure in Figure 3), which is consistent with the Insight data. The thermoelastic properties (such as α, $K_T$, γ) are adopted directly from the pressure-temperature-dependent results from Section 3.2. Thermal conductivity is directly set as a constant (22 W/m/K). The CMB heat flow is a parameterized model [5] with $Q_{cmb}(t)=Q_0 exp(-t/\tau)$ ($\tau$ = 0.69 Ga), which is restricted by the cessation of the dynamo in the early 0.5 Ga. The $Q_0$=0.9 TW is appropriate for the cessation of the dynamo, which is consistent with Ref. [11]. By the above selected physical parameters, the iron snow exists at present and the presence of iron snow cannot reactivate the Martian dynamo, in the subfigure of Figure 3.

The thermal evolution of the core is sensitive to the basic physical property, such as current temperature at CMB $T_{cmb}(t=0)$, the radius of core $R_c$, the density of core $\rho$, CMB heat flux $Q_{cmb}(t)$, and thermal conductivity $k$. A systematic calculation of the parameter space was explored in Figure 3. The thickness of the snow zone ($r_{cmb}$-$R_s$) and cessation time of the dynamo is selected to explore parameter uncertainty about $T_{cmb}$, $R_c$, $\rho$, $Q_0$, and $k$. The thickness of the snow zone is sensitive to the present CMB temperature, as the location of the snow zone is all below the melting temperature. The cessation time of the dynamo is sensitive to the thermal conductivity and CMB heat flux, which directly affect the heat transport of the core.

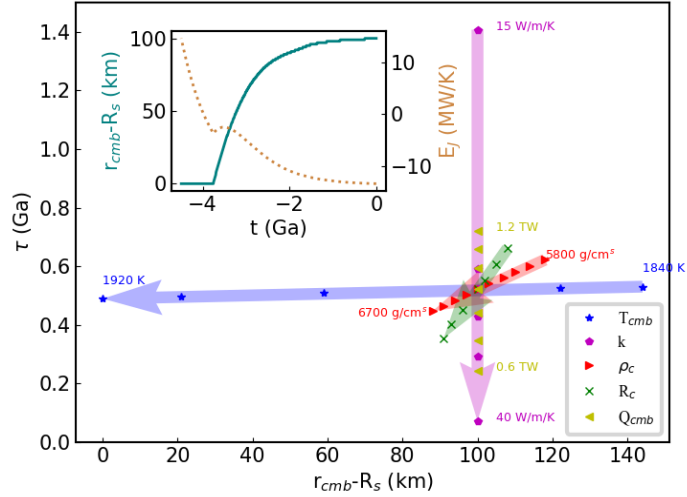

Figure 3 Phase-diagram of the iron snow zone thickness and cessation time of dynamo with different parameters. The parameter variation includes $T_{cmb}$, $R_c$, $\rho$, $Q_0$, and $k$. The central point with initial condition: $T_{cmb}$ = 1860 K, $P_{cmb}$ = 18 GPa, $R_c$ = 1830 km, $\rho_c$ = 6300 kg/cm3, $Q_{cmb}$(t=-4.5) = 0.9 TW, and $k$ = 22 W/m/K. The subfigure is the evolution of the radius of the snow zone ($R_s$) and Ohmic dissipation entropy ($E_J$) in the Martian core.

## 4. Conclusions

To study the thermoelastic properties and thermal evolution of the Martian core, ab initio calculations are adopted to calculate the equation of state and thermal conductivity of liquid $Fe_{0.75}S_{0.25}$. The results show that liquid $Fe_{0.75}S_{0.25}$ under the Martian core condition is thoroughly ferromagnetic. The sound velocity ranges from 4.21 to 5.19 km/s. The thermal conductivity is 21~23 W/m/K in the spin-polarized calculation but is 40~55 W/m/K in the spin-non-polarized calculation. In the thermal evolution calculation, the possible present evolution mechanism of the Martian core is simply secular cooling or iron snow model. The parameter uncertainty effect on the cessation time of dynamo and zone of iron snow is explored.

**Acknowledgement**

This work was support by the National Natural Science Foundation of China (NSFC) [grant numbers 11975058, 11775031 and 11625415], and the fund of Key Laboratory of Computational Physics [grant number 6142A05RW202103]. We thanked for the fund support from Laboratory of Computational Physics in Institute of Applied Physics and Computational Mathematics.

**References**

1    Stevenson, D. J. Mars' core and magnetism. *Nature* **412**, 214-219, doi:10.1038/35084155 (2001).

2    Smrekar, S. E. *et al.* Pre-mission InSights on the Interior of Mars. *Space Science Reviews* **215**, 3, doi:10.1007/s11214-018-0563-9 (2018).

3    Stähler, S. C. *et al.* Seismic detection of the martian core. *Science* **373**, 443-448,

doi:doi:10.1126/science.abi7730 (2021).

4  Knapmeyer-Endrun, B. *et al.* Thickness and structure of the martian crust from InSight seismic data. *Science* **373**, 438-443, doi:doi:10.1126/science.abf8966 (2021).

5  Williams, J.-P. & Nimmo, F. Thermal evolution of the Martian core: Implications for an early dynamo. *Geology* **32**, 97-100, doi:10.1130/G19975.1 (2004).

6  Gilfoy, F. & Li, J. Thermal state and solidification regime of the martian core: Insights from the melting behavior of FeNi-S at 20 GPa. *Earth and Planetary Science Letters* **541**, 116285, doi:https://doi.org/10.1016/j.epsl.2020.116285 (2020).

7  Stewart, A. J., Schmidt, M. W., Westrenen, W. v. & Liebske, C. Mars: A New Core-Crystallization Regime. *Science* **316**, 1323-1325, doi:doi:10.1126/science.1140549 (2007).

8  Davies, C. J. & Pommier, A. Iron snow in the Martian core? *Earth and Planetary Science Letters* **481**, 189-200, doi:10.1016/j.epsl.2017.10.026 (2018).

9  Greenwood, S., Davies, C. J. & Pommier, A. Influence of Thermal Stratification on the Structure and Evolution of the Martian Core. *Geophysical Research Letters* **48**, e2021GL095198, doi:https://doi.org/10.1029/2021GL095198 (2021).

10 Yokoo, S., Hirose, K., Tagawa, S., Morard, G. & Ohishi, Y. Stratification in planetary cores by liquid immiscibility in Fe-S-H. *Nature Communications* **13**, 644, doi:10.1038/s41467-022-28274-z (2022).

11 Hemingway, D. J. & Driscoll, P. E. History and Future of the Martian Dynamo and Implications of a Hypothetical Solid Inner Core. *Journal of Geophysical Research: Planets* **126**, e2020JE006663, doi:https://doi.org/10.1029/2020JE006663 (2021).

12 Rivoldini, A., Van Hoolst, T., Verhoeven, O., Mocquet, A. & Dehant, V. Geodesy constraints on the interior structure and composition of Mars. *Icarus* **213**, 451-472, doi:https://doi.org/10.1016/j.icarus.2011.03.024 (2011).

13 Khan, A. *et al.* A Geophysical Perspective on the Bulk Composition of Mars. *Journal of Geophysical Research: Planets* **123**, 575-611, doi:https://doi.org/10.1002/2017JE005371 (2018).

14 Nishida, K. *et al.* Effect of sulfur on sound velocity of liquid iron under Martian core conditions. *Nature Communications* **11**, 1954, doi:10.1038/s41467-020-15755-2 (2020).

15 Williams, Q. The thermal conductivity of Earth's core: a key geophysical parameter's constraints and uncertainties. *Annual Review of Earth Planetary Sciences* **46**, 47-66 (2018).

16 Suehiro, S., Ohta, K., Hirose, K., Morard, G. & Ohishi, Y. The influence of sulfur on the electrical resistivity of hcp iron: Implications for the core conductivity of Mars and Earth. *Geophysical Research Letters* **44**, 8254-8259, doi:https://doi.org/10.1002/2017GL074021 (2017).

17 Ezenwa, I. C. & Yoshino, T. Martian core heat flux: Electrical resistivity and thermal conductivity of liquid Fe at Martian core P-T conditions. *Icarus* **360**, doi:10.1016/j.icarus.2021.114367 (2021).

18 Hsieh, W.-P. *et al.* Low thermal conductivity of iron-silicon alloys at Earth's core conditions with implications for the geodynamo. *Nature Communications* **11**, 3332, doi:10.1038/s41467-020-17106-7 (2020).

19 Birch, F. Elasticity and constitution of the Earth's interior. *Journal of Geophysical Research* **57**, 227-286 (1952).

20 Chester, G. & Thellung, A. The law of Wiedemann and Franz. *Proceedings of the Physical*

*Society* **77**, 1005 (1961).
21 Kresse, G. & Hafner, J. Ab initio molecular dynamics for liquid metals. *Physical Review B* **47**, 558 (1993).
22 Kresse, G. & Furthmüller, J. Efficient iterative schemes for ab initio total-energy calculations using a plane-wave basis set. *Physical Review B* **54**, 11169 (1996).
23 Blöchl, P. E. Projector augmented-wave method. *Physical Review B* **50**, 17953 (1994).
24 Kresse, G. & Joubert, D. From ultrasoft pseudopotentials to the projector augmented-wave method. *Physical Review B* **59**, 1758 (1999).
25 Perdew, J. P., Burke, K. & Ernzerhof, M. Generalized gradient approximation made simple. *Physical Review Letters* **77**, 3865 (1996).
26 Cococcioni, M. & De Gironcoli, S. Linear response approach to the calculation of the effective interaction parameters in the LDA+ U method. *Physical Review B* **71**, 035105 (2005).
27 Kulik, H. J., Cococcioni, M., Scherlis, D. A. & Marzari, N. Density functional theory in transition-metal chemistry: A self-consistent Hubbard U approach. *Physical Review Letters* **97**, 103001 (2006).
28 Lin, J.-F. *et al.* Spin transition zone in Earth's lower mantle. *Science* **317**, 1740-1743 (2007).
29 Lin, J.-F. *et al.* Magnetic transition and sound velocities of Fe3S at high pressure: implications for Earth and planetary cores. *Earth and Planetary Science Letters* **226**, 33-40, doi:https://doi.org/10.1016/j.epsl.2004.07.018 (2004).